\documentclass[%
 reprint,
superscriptaddress,
 amsmath,amssymb,
 aps,
pra,
]{revtex4-2}

\usepackage{microtype}
\usepackage{graphicx}
\usepackage{dcolumn}
\usepackage{bm}
\usepackage[colorlinks=true,bookmarks=false,citecolor=blue,urlcolor=blue]{hyperref}
\usepackage{xcolor}
\definecolor{darkblue}{rgb}{0,0.3,0.7}
\definecolor{teal}{RGB}{26,157,150}
\definecolor{softpurple}{RGB}{135,110,160}

\newcommand\imag{\mathrm{i}}
\usepackage{verbatim,upgreek,braket}

\hypersetup{
colorlinks=true,
linkcolor=darkblue,
filecolor=blue,
citecolor=darkblue,  
urlcolor=darkblue,
}

\usepackage[capitalize]{cleveref}
\crefname{section}{Sec.}{Secs.}
\Crefname{section}{Section}{Sections}

\crefrangelabelformat{section}{#3#1#4--#5\crefstripprefix{#1}{#2}#6}
\crefrangelabelformat{figure}{#3#1#4--#5\crefstripprefix{#1}{#2}#6}
\crefrangelabelformat{equation}{(#3#1#4--#5\crefstripprefix{#1}{#2}#6)}

\crefmultiformat{equation}%
{\edef\crefstripprefixinfo{#1}Eqs.~(#2#1#3}%
{,#2\crefstripprefix{\crefstripprefixinfo}{#1}#3)}%
{,#2\crefstripprefix{\crefstripprefixinfo}{#1}#3}%
{,#2\crefstripprefix{\crefstripprefixinfo}{#1}#3)}

\begin{document}

\preprint{APS/123-QED}

\title{Frequency-bin-encoded quantum secret sharing with fast basis switching}

\author{Chen-You Su}
\altaffiliation{Equal contribution.}
\affiliation{Elmore Family School of Electrical and Computer Engineering and Purdue Quantum Science and Engineering Institute, Purdue University, West Lafayette, IN 47907, USA}
\affiliation{Department of Physics and Astronomy, Purdue University, West Lafayette, IN 47907, USA}

\author{Suparna Seshadri}
\altaffiliation{Equal contribution.}
\affiliation{Elmore Family School of Electrical and Computer Engineering and Purdue Quantum Science and Engineering Institute, Purdue University, West Lafayette, IN 47907, USA}
\affiliation{Aliro Technologies, Inc.,
Brighton, MA 02135 USA}

\author{Hsuan-Hao Lu}
\affiliation{Elmore Family School of Electrical and Computer Engineering and Purdue Quantum Science and Engineering Institute, Purdue University, West Lafayette, IN 47907, USA}

\author{Jason~D.~McKinney}
\affiliation{Elmore Family School of Electrical and Computer Engineering and Purdue Quantum Science and Engineering Institute, Purdue University, West Lafayette, IN 47907, USA}

\author{Andrew M. Weiner}
\altaffiliation{Deceased.}
\affiliation{Elmore Family School of Electrical and Computer Engineering and Purdue Quantum Science and Engineering Institute, Purdue University, West Lafayette, IN 47907, USA}

\author{Joseph M. Lukens}
\email{jlukens@purdue.edu}
\affiliation{Elmore Family School of Electrical and Computer Engineering and Purdue Quantum Science and Engineering Institute, Purdue University, West Lafayette, IN 47907, USA}
\affiliation{Quantum Information Science Section, Oak Ridge National Laboratory, Oak Ridge, Tennessee 37831, USA}
\date{\today}

\begin{abstract}
Frequency-bin encoding offers unique opportunities for fiber-native quantum information processing, yet its full potential has been hampered by challenges in rapid state control and protocol scaling.
Leveraging high-speed microwave processing, we overcome several extant hurdles and unlock new fundamental capabilities in frequency-based quantum communications. First, we demonstrate three-party quantum secret sharing, representing, to our knowledge, the first multiparty ($N\geq 3$ users) quantum communication protocol implemented entirely with frequency encoding. Second, we demonstrate active, pseudorandom switching of frequency-bin quantum states and measurement bases at 20 MHz symbol rates, enabling real-time protocol operation without relying on hardware-intensive passive basis-selection schemes. Even under fast switching, we preserve high-contrast three-party correlations ($\geq 0.91$) and low quantum bit error rates ($\leq 4.5\%$). Our work extends frequency-bin quantum communication beyond static, two-party protocols, opening a pathway toward scalable, reconfigurable, and real-time spectrally encoded quantum networking.
\end{abstract}

\maketitle


\section{\label{sec:level0}Introduction}
Emerging quantum networks naturally rely on the existing lightwave infrastructure~\cite{kimble2008quantum,wehner2018quantum,Lukens2025}, making the photonic degree of freedom for encoding critical for both stability and parallelizability. 
Among the many options available---such as path, orbital angular momentum (OAM), and time bins~\cite{flamini2019photonic,erhard2020advances}---frequency-bin encoding~\cite{lukens2016frequency,lu2023frequency} is distinctive in requiring less demanding interferometric stabilization than time-bin schemes~\cite{islam2017provably,yu2025quantum} while supporting easily multiplexable orthogonal modes within a conventional single-mode fiber, unlike path~\cite{ding2017high,zahidy2024practical} and OAM~\cite{xavier2020quantum,wang2022quantum}.
Frequency-bin quantum states can also be manipulated using electro-optic modulation and programmable spectral filters inherited from mature telecommunication technologies~\cite{lukens2016frequency,lu2023frequency}. 
These tools support parallel processing across frequency channels and have enabled a range of frequency-bin quantum operations, including frequency beamsplitters~\cite{lu2018electro}, a two-qubit entangling gate~\cite{lu2019controlled}, and universal state control~\cite{lu2020fully}.

Spurred by these advances---as well as an older, independent thread in quantum key distribution (QKD) with time-frequency encodings~\cite{merolla1999single,merolla1999quantum,bloch2007frequency,ali2007large,nunn2013large}---several entanglement-based frequency-bin QKD demonstrations have emerged in the last two years. In one approach~\cite{tagliavacche2025frequency}, fiber couplers randomly direct photons toward parallel electro-optic-based analyzers, while another~\cite{khodadad2025frequency} combines electro-optic modulation with frequency-to-time mapping to reduce the number of detectors; others have begun exploring higher-dimensional qutrit QKD with single tunable receivers~\cite{crisan2026multi}. 
Nevertheless, well-established quantum communication capabilities available in more traditional encoding formats---such as fast state selection and multiparty protocols---have yet to be demonstrated in frequency bins.

In their archetypal form, quantum communication protocols based on conjugate encodings typically assume fast and random basis selection, where ``fast'' in this context means ``faster than the detection rate''~\cite{Sasaki2016}, for otherwise adjacent events would be correlated in their measurement bases.
Although fast basis selection can be replaced by passive splitters
in entanglement-based QKD~\cite{ekert1991quantum,bennett1992quantum,jennewein2000quantum,ma2007quantum}, 
passive selection roughly doubles the number of components at each receiver, introduces nontrivial nuances in the security proofs when bias is present~\cite{Kawakami2025, Ivchenko2026}, and restricts the protocols that can be realized in practice---excluding, e.g., loophole-free Bell tests~\cite{Hensen2015, Giustina2015, Shalm2015} and prepare-and-measure QKD~\cite{wiesner1983conjugate, Bennett1984BB84,  gisin2002quantum}. 
In addition, ideal quantum networks should support secure multiparty applications beyond two-party QKD alone. One such protocol, quantum secret sharing (QSS), requires a set number of participants to cooperate to recover a secret that is inaccessible to any smaller subset~\cite{Blakley1979Safeguarding,shamir1979share,hillery1999quantum,xiao2004efficient}. Previous QSS experiments have employed energy-time~\cite{tittel2001experimental}, polarization~\cite{chen2005experimental,schmid2005experimental,williams2019quantum,cabrejo2025frequency}, and phase encodings~\cite{bogdanski2008experimental,hai2013experimental}, but so far no multiparty frequency-bin-encoded protocol has been realized.

In this work, we address these missing capabilities and demonstrate three-party QSS encoded entirely in frequency bins with high-speed active electrical switching. Building on a two-photon polarization-based QSS protocol in which the source functions as the third participant~\cite{williams2019quantum},
we leverage the equivalence between linear spectral phase and temporal delay to transfer state preparation and basis selection from the frequency domain to the time domain, where radio-frequency (RF) phase shifters enable fast state and measurement reconfiguration. 
Compared with the ms-scale switching times of liquid-crystal-on-silicon (LCoS) pulse shapers conventionally used for frequency-bin state control~\cite{seshadri2022complete}, the ns-scale speed of analog phase shifters accelerates basis selection to the regime of real-time operation.
Demonstrating periodic and pseudorandom patterns at 20~MHz symbol rates, we obtain three-party QSS correlations with quantum bit error rates (QBERs) under 5\% in all cases---comfortably below the 11\% positive-key threshold~\cite{shor2000simple,koashi2003secure,ma2007quantum,williams2019quantum}.
Significantly, this rate is already approaching the maximum useful switching speed set by the detector dead time, suggesting broad opportunities for fast, multiparty quantum communications in frequency-bin encoding at the limits of photonic hardware.

\section{\label{sec:level1}Background}
\begin{figure}[t]
    \centering    \includegraphics[width=\columnwidth]{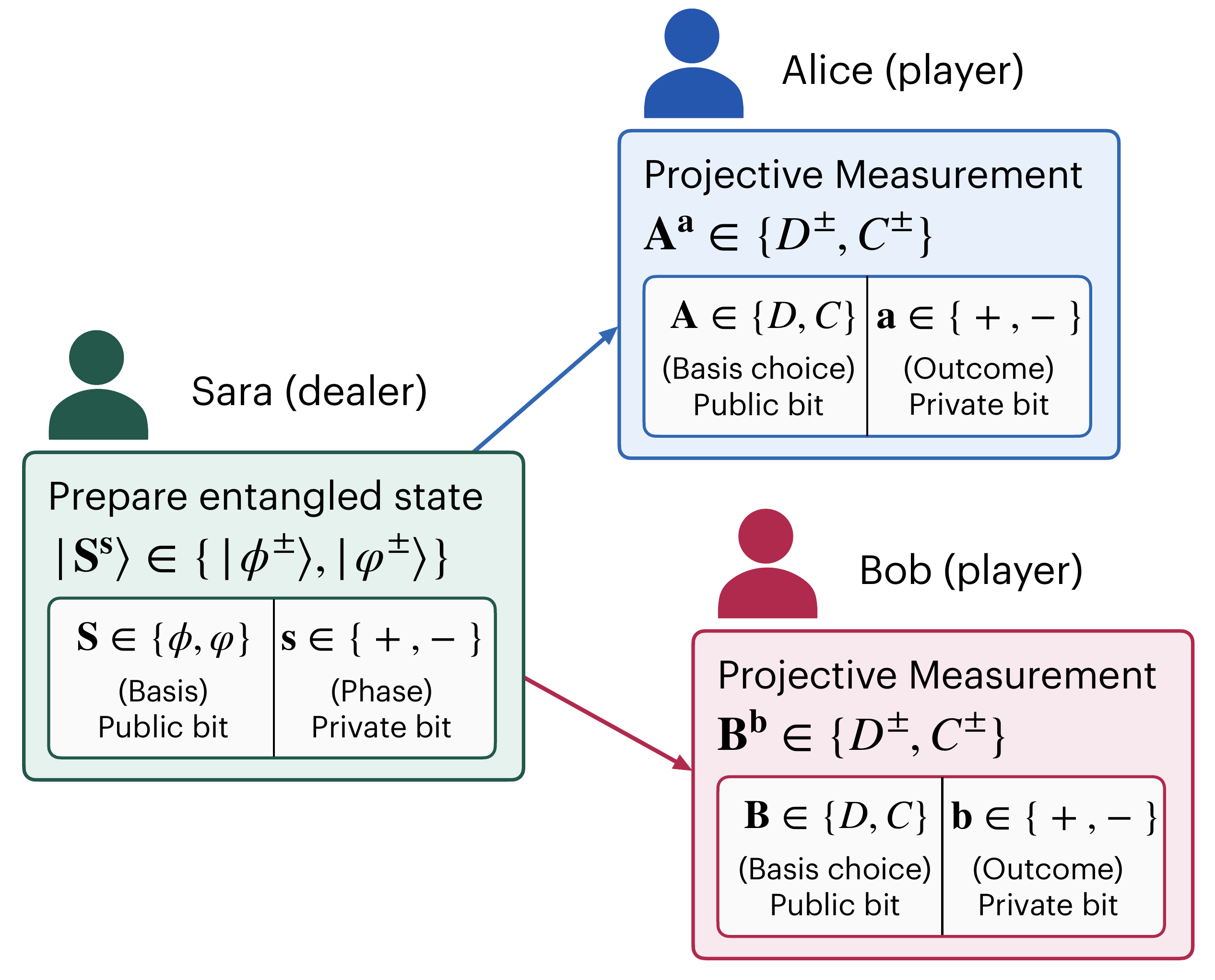}
    \caption{Schematic of the three-party QSS protocol. Sara prepares one of four entangled states, $\ket{\mathbf{S^s}}\in\{\ket{\phi^\pm},\ket{\varphi^\pm}\}$, specified by her public bit $\mathbf{S}\in\{\phi,\varphi\}$ and private bit $\mathbf{s}\in\{+,-\}$. The two photons are then distributed to Alice and Bob, who independently perform projective measurements in either the diagonal ($D$) or circular ($C$) basis, with $\mathbf{A},\mathbf{B}\in\{D,C\}$ denoting their public basis choices and $\mathbf{a},\mathbf{b}\in\{+,-\}$ their private measurement outcomes. The public bits are disclosed during a later post-processing stage, whereas the private bits remain secret: those held by Alice and Bob serve as partial keys, while Sara’s private bit encodes the secret.}
    \label{fig:protocol}
\end{figure}
Many cryptographic applications require cooperation among multiple parties, including remote voting, online auctions, secure money transfer, and secure multiparty computation, for which QKD alone does not directly provide the required multipartite access structure. QSS is a key cryptographic primitive for such settings. The concept of classical secret sharing was independently introduced by Shamir~\cite{shamir1979share} and Blakley~\cite{Blakley1979Safeguarding} in 1979 and later extended to quantum settings~\cite{hillery1999quantum}; throughout this work, QSS refers specifically to the use of quantum resources to share a \textit{classical} secret, rather than an unknown quantum state.

Within the general QSS framework, a dealer distributes a quantum state among $\mathcal{N}$ players such that the secret can be reconstructed only through the cooperation of at least $k$ players, defining a $(k,\mathcal{N})$-threshold scheme. QSS fundamentally relies on quantum correlations shared among the participating parties. In the original Hillery-Bužek-Berthiaume protocol~\cite{hillery1999quantum}, these correlations are established by an $(\mathcal{N}+1)$-photon GHZ state shared by one dealer and $\mathcal{N}$ players. However, photonic GHZ states are experimentally demanding to generate, particularly so for frequency bins where GHZ states remain theoretical proposals~\cite{banic2024integrated} without experimental realization. Thankfully, subsequent analysis~\cite{karlsson1999quantum} showed that QSS can instead be implemented using two-photon entanglement. Building on this biphoton framework, our work adopts the modified QSS protocol of Ref.~\cite{williams2019quantum}, in which a source of entangled photons, two receivers, and intermediate parties applying unitary rotations provide $\mathcal{N}+1\geq 3$ users for the protocol.

We experimentally focus on a three-party instantiation of the QSS protocol, involving one dealer (Sara) and two players (Alice and Bob)~\footnote{The roles of the three parties are interchangeable; the particular assignment used here is adopted solely for clarity.}, as illustrated in Fig.~\ref{fig:protocol}. This configuration corresponds to the $(k,\mathcal{N})=(2,2)$ case, in which both players must jointly use their private bits to recover the dealer's  secret. To implement this protocol, Sara prepares one of the following four states:
\begin{equation}
\begin{aligned}
\ket{\phi^\pm}
&=
\frac{1}{\sqrt{2}}
\left(
\ket{A_0 B_0}
\pm
\ket{A_1 B_1}
\right),\\
\ket{\varphi^\pm}
&=
\frac{1}{\sqrt{2}}
\left(
\ket{A_0B_0}
\pm \imag\ket{A_1 B_1}
\right).
\end{aligned}
\label{eq:states}
\end{equation}
Here, $\ket{A_n}$ and $\ket{B_m}$ denote the single-photon Fock states occupying the $n^{\mathrm{th}}$ and $m^{\mathrm{th}}$ frequency bins of the signal and idler photons routed to Alice and Bob, respectively. Sara’s public bit $\mathbf{S}$ specifies the basis choice, $\mathbf{S}\in\{\phi,\varphi\}$, whereas her private bit $\mathbf{s}$ specifies the relative-phase choice within that basis, $\mathbf{s}\in\{+,-\}$, which encodes the secret.

Upon receiving their photons, Alice and Bob independently measure in either the diagonal ($D$) or circular ($C$) basis, with corresponding eigenstates:
\begin{equation}
\begin{aligned}
\ket{D^\pm}_A
&=
\frac{1}{\sqrt{2}}
\left(\ket{A_0}\pm\ket{A_1}\right),\\
\ket{C^\pm}_A
&=
\frac{1}{\sqrt{2}}
\left(\ket{A_0}\pm \imag\ket{A_1}\right),\\
\ket{D^\pm}_B
&=
\frac{1}{\sqrt{2}}
\left(\ket{B_0}\pm\ket{B_1}\right),\\
\ket{C^\pm}_B
&=
\frac{1}{\sqrt{2}}
\left(\ket{B_0}\pm \imag\ket{B_1}\right).
\end{aligned}
\label{eq:meas}
\end{equation}
Alice's public bit $\mathbf{A}\in\{D,C\}$ specifies her measurement basis, while her private bit $\mathbf{a}\in\{+,-\}$ records the corresponding measurement outcome. The same notation applies to Bob, with $\mathbf{B}\in\{D,C\}$ denoting his basis choice and $\mathbf{b}\in\{+,-\}$ his measurement outcome.

Expressing \cref{eq:states} in terms of the measurement bases defined in \cref{eq:meas} reveals the correlations required for QSS:
\begin{equation}
\begin{aligned}
\ket{\phi^\pm}
&=
\frac{1}{\sqrt{2}}
\left(
\ket{D^+D^\pm}
+
\ket{D^-D^\mp}
\right)\\
&=
\frac{1}{\sqrt{2}}
\left(
\ket{C^+C^\mp}
+
\ket{C^-C^\pm}
\right),\\
\ket{\varphi^\pm}
&=
\frac{1}{\sqrt{2}}
\left(
\ket{D^+C^\pm}
+
\ket{D^-C^\mp}
\right)\\
&=
\frac{1}{\sqrt{2}}
\left(
\ket{C^+D^\pm}
+
\ket{C^-D^\mp}
\right).
\end{aligned}
\label{eq:correlations}
\end{equation}
Ideally, provided that Sara selects from $\mathbf{s} \in \{+, -\}$ with equal probability, the valid basis combinations---when Alice and Bob choose the same basis for $\ket{\phi^\pm}$ or different basis for $\ket{\varphi^\pm}$---yield private bits $(\mathbf{s},\mathbf{a},\mathbf{b})$ that are individually random and pairwise uncorrelated, yet collectively obey a deterministic three-party correlation. To illustrate, consider the public bit choices $(\mathbf{S},\mathbf{A},\mathbf{B})=(\phi,D,D)$. If Alice obtains $\mathbf{a}=+$, \cref{eq:correlations} leaves two equally probable joint triplets,
\[
(\mathbf{s},\mathbf{a},\mathbf{b})=(+,+,+)
\quad\text{or}\quad
(-,+,-).
\]
Alice's outcome alone therefore provides no information about Sara's
private bit. Once Bob shares his outcome, however, Sara's bit is
uniquely determined: if Bob obtains $+$, then Sara's bit is $+$; if Bob obtains $-$, then Sara's is $-$. 

Note that the prepared state and measurement basis defined in
\cref{eq:states,eq:meas} are mutually
unbiased, satisfying
$\left|\braket{\varphi^{i}|\phi^{j}}\right|^2
=\left|\braket{C^{i}|D^{j}}\right|^2=\frac{1}{2}$
for $i,j\in\{+,-\}$. Thus, a measurement that perfectly distinguishes the states in one basis yields completely random outcomes for the other. Security against an external eavesdropper follows the similar complementarity-based principle as QKD; a complete description of the protocol and its security analysis is provided in Ref.~\cite{williams2019quantum}. 
It is also worth noting that while Alice and Bob can alternatively select their bases passively with a beamsplitter and two separately operating $C$ and $D$ apparatuses, practical operation requires Sara to \emph{actively} select the prepared state so that her public and private bits are well defined, highlighting the importance of the high-speed active switching demonstrated here.

\section{\label{sec:Exp}Experiment}

Since all of Sara's four states [\cref{eq:states}] and Alice and Bob's measurement projections [\cref{eq:meas}] differ solely by spectral phase, it is in principle possible to perform all switching in the frequency domain: namely, by adjusting the phase applied by line-by-line pulse shapers at Sara, Alice, and Bob. However, standard LCoS pulse shaping technology typically achieves update times on the order of several hundred ms~\cite{zheng2019microwave, wang200120}---with $\lesssim$1~ms possible with appropriate engineering~\cite{Pivnenko2021}---far too slow for the MHz symbol rates desired in quantum communications.
Although less common electrically and electro-optically reconfigurable line-by-line shapers could reach such rates~\cite{ni2023high, akbulut2010dynamic}, we opt for a more direct approach with electro-optic intensity and phase modulators (EOIMs and EOPMs). 
By the Fourier equivalence between temporal delay and linear spectral phase, the relative phase between any two frequency bins can be set by an appropriately tuned delay~\cite{weiner2011ultrafast,seshadri2022nonlocal}. Although not extensible to independent control of more than two arbitrary phases---since these no longer define a line---for the qubit cases here delay-based control is fully sufficient. Indeed, in prior frequency-bin experiments this relationship has been used to simplify gate design~\cite{lu2020fully} and reduce the number of components in projective measurements~\cite{lu2023characterization}.

\subsection{Fast switching}

\begin{figure*}[t]
    \centering    \includegraphics[width=\textwidth]{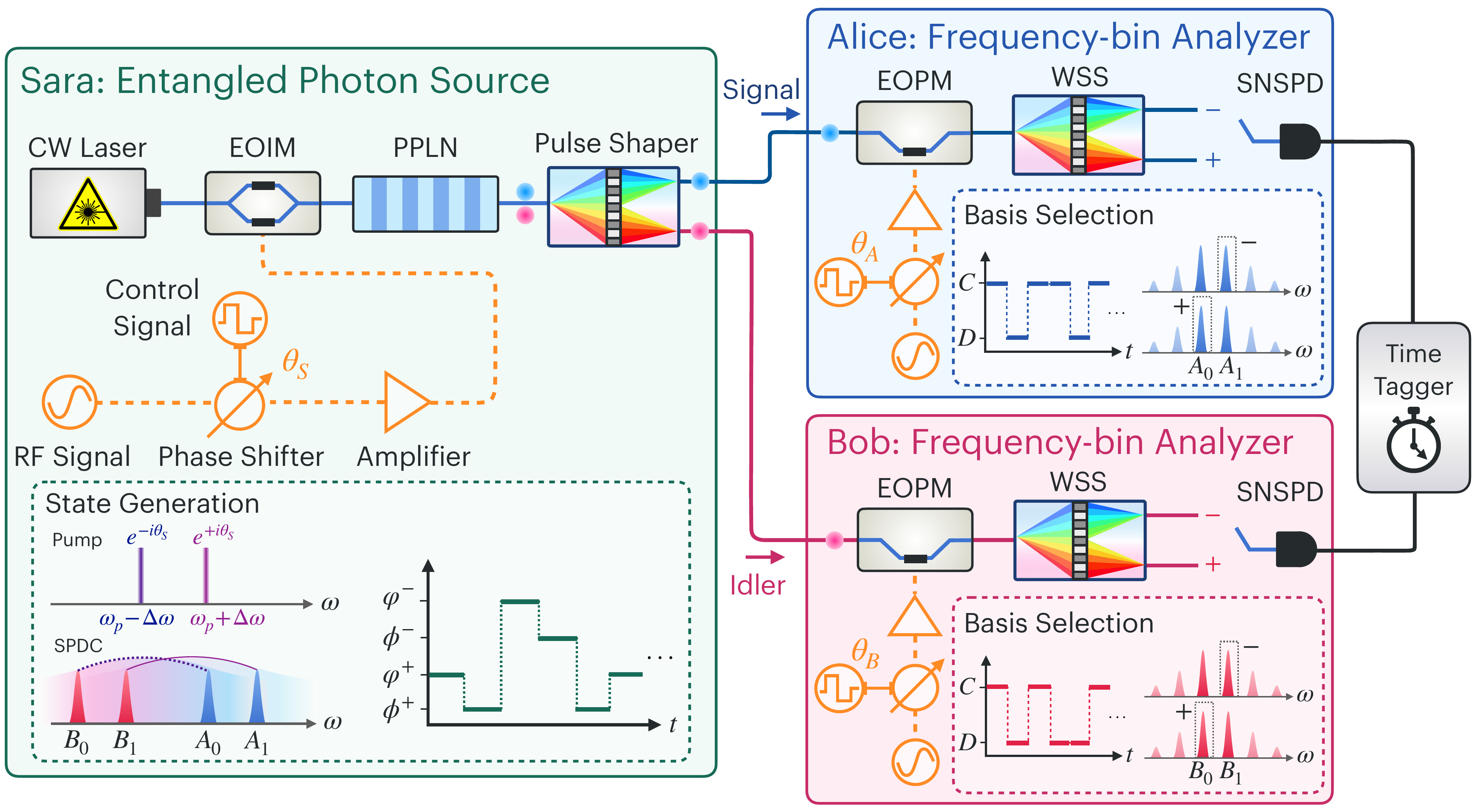}
    \caption{
    Simplified experimental setup for three-party frequency-bin QSS. Sara generates frequency-bin-entangled signal-idler photon pairs by dual-line pumping a PPLN waveguide and uses a pulse shaper to select and route the signal and idler frequency bins to Alice and Bob, respectively. Fast control of the RF phase $\theta_S$ applied to the source EOIM enables preparation of the four states $\phi^{+}$, $\phi^{-}$, $\varphi^{+}$, and $\varphi^{-}$ (state-generation inset). Alice and Bob analyze their photons using EOPMs, by switching their RF phases ($\theta_A$, $\theta_B$) and selecting between the diagonal ($D$) and circular ($C$) measurement bases (basis-selection insets). The EOPMs map the two projection outcomes, $+$ and $-$, onto distinct frequency bins, which are then selectively routed by the WSSs to the respective SNSPDs at Alice and Bob. The detection events are recorded by an event timer for coincidence analysis.
    Abbreviations: CW, continuous wave; EOIM, electro-optic intensity modulator; PPLN, periodically poled lithium niobate; EOPM, electro-optic phase modulator; WSS, wavelength-selective switch; SNSPD, superconducting nanowire single-photon detector.}
    \label{fig:setup}
\end{figure*}

The experimental setup for our frequency-bin-encoded QSS implementation is shown in Fig.~\ref{fig:setup}. 
Following the approach proposed in Ref.~\cite{seshadri2022complete}, Sara prepares the entangled states defined in Eq.~\eqref{eq:states}. Although perhaps surprising at first glance, we concentrate on the more complicated \emph{positive-frequency-correlated} Bell states, which require bichromatic pumping for generation~\cite{seshadri2022complete, Clementi2023}. This choice is motivated by speed and efficiency: because the dual-line pump is already created by an EOIM, it is possible to adjust the state phase rapidly in the RF domain prior to SPDC. In contrast, the standard negatively correlated Bell state obtained by a single-line pump can only be modulated post-generation, which would require another downstream electro-optic modulator for MHz-rate switching.

Consider a continuous-wave (CW) pump field at frequency $\omega_p$ passing through an EOIM biased at its transmission null and driven by the RF waveform $V_\mathrm{RF}\cos(\Delta\omega t-\theta_S)$. The complex field at the output can then be written as
\begin{equation}
E_{\mathrm{pump}}(t)
\propto
e^{-\imag\omega_p t}
\sin\!\left[\delta\cos(\Delta\omega t-\theta_S)\right],
\end{equation}
where $\delta=\frac{\pi V_\mathrm{RF}}{2V_\pi}$ is the modulation depth set by the RF amplitude $V_\mathrm{RF}$ relative to the half-wave voltage $V_\pi$ of the EOIM. By the Jacobi-Anger expansion~\cite{Capmany2011review}, the two first-order sidebands can be written as
\begin{equation}
\label{eq:JacobiAnger}
E_{\mathrm{pump}}^{(\pm1)}(t)
\propto
J_1(\delta)\,e^{-\imag(\omega_p\pm\Delta\omega)t}\,e^{\pm \imag\theta_S},
\end{equation}
where $J_\ell$ denotes the $\ell^\text{th}$-order Bessel function of the first kind.
At null bias, only odd-order sidebands are generated, suppressing the carrier; at our operating point, the third- and higher-order sidebands are negligible.
As illustrated in the state-generation inset of Fig.~\ref{fig:setup}, the two mutually coherent pump lines at $\omega_p\pm\Delta\omega$, carrying phases $e^{\pm\imag\theta_S}$, drive separate SPDC processes that generate photon pairs centered at half their respective frequencies. The pulse shaper carves the corresponding frequency bins $\{A_1,B_1\}$ and $\{A_0,B_0\}$, yielding the postselected state
\begin{equation}
\ket{\Psi(\theta_S)} = \frac{1}{\sqrt{2}}
\left(\ket{A_0B_0} + e^{2\imag\theta_S}\ket{A_1B_1}\right).
\label{eq:State_S}
\end{equation}
Sara can therefore prepare any of the four states in \cref{eq:states} solely by setting the \emph{RF phase} of her EOIM drive: $\theta_S\in\{0,\frac{\pi}{4},\frac{\pi}{2},\frac{3\pi}{4}\}$ for $\{\ket{\phi^+},\ket{\varphi^+},\ket{\phi^-},\ket{\varphi^-}\}$, respectively.

Alice and Bob select their measurement bases $\mathbf{A},\mathbf{B}\in\{D,C\}$ in the same way, by setting the RF phase $\theta_X$ ($X\in\{A,B\}$) of their EOPM drives (basis-selection insets of Fig.~\ref{fig:setup}). Let $\hat{a}_{X_k}$ and $\hat{b}_{X_k}$ denote the annihilation operators for frequency bin $X_k$ ($k\in\{0,1\}$) immediately before and after the EOPM.
For phase modulation of the form $\exp\!\left[\imag m\sin(\Delta\omega t-\theta_X)\right]$, the output annihilation operators are given by~\cite{lu2020fully,seshadri2022complete, Capmany2011review}
\begin{equation}
    \hat{b}_{X_k}
    =
    \sum_\ell
    J_\ell(m)\,e^{-\imag\ell\theta_X}\,\hat{a}_{X_{k+\ell}}.
\end{equation}
At $m = 1.435~\mathrm{rad}$, $J_1(m)=J_0(m)=-J_{-1}(m)=0.548$. Therefore, restricted to the computational space, the transformation becomes
\begin{equation}
\begin{pmatrix}
\hat b_{X_0}\\
\hat b_{X_1}
\end{pmatrix}
=
\frac{\eta}{\sqrt{2}}
\begin{pmatrix}
1 & e^{-\imag\theta_X}\\
-e^{\imag\theta_X} & 1
\end{pmatrix}
\begin{pmatrix}
\hat a_{X_0}\\
\hat a_{X_1}
\end{pmatrix}.
\label{eq:eopm_gate}
\end{equation}
The prefactor $\eta$ accounts for the probabilistic
nature of the operation: photons scattered outside the computational bins are discarded, giving a success probability $\mathcal{P}\equiv\eta^2=2J_1^2(m)\simeq0.60$.
Setting $\theta_X=0$ ($\frac{\pi}{2}$) thus implements a measurement in the $D$ ($C$) basis, with the $+$ and $-$ outcomes routed to output bins $X_0$ and $X_1$, respectively.
This is analogous to the spatial separation of spin eigenstates in a Stern-Gerlach apparatus, or of polarization projections by a waveplate and polarizing beamsplitter.

Consequently, the RF phases $\theta_S\in\left\{0,\frac{\pi}{4},\frac{\pi}{2},\frac{3\pi}{4}\right\}$ and $\theta_A,\theta_B\in\left\{0,\frac{\pi}{2}\right\}$ completely specify all unique QSS settings---the public bits of all three parties and Sara's private bit. Alice's and Bob's private bits are then revealed by the output bins in which their photons are detected. The corresponding coincidence probability is
\begin{equation}
    P_{{A}_{j}{B}_{k}}(\theta_S,\theta_A,\theta_B)
    =
    \left|
        \braket{\mathrm{vac}|
        \hat{b}_{{A}_{j}}(\theta_A)
        \hat{b}_{{B}_{k}}(\theta_B)|
        \Psi(\theta_S)}
    \right|^{2},
    \label{eq:CoinProb}
\end{equation}
for $j,k\in\{0,1\}$.

\subsection{Hardware details}
Our pump is derived from a CW laser (\textit{Toptica}) operating at $780.04~\mathrm{nm}$ ($\omega_p/2\pi=384.33~\mathrm{THz}$), which is sent through an EOIM (\textit{Exail})
biased at its transmission null and driven at an RF frequency of $\Delta\omega/2\pi =25~\mathrm{GHz}$ with modulation depth $\delta\approx1.03$. Carrier-suppressed modulation generates two dominant pump lines at $\omega_p\pm\Delta\omega$, with the residual carrier and the third-order sidebands suppressed by approximately 18 and 26~dB. The two pump lines, with a combined optical power of $0.8~\mathrm{mW}$, are directed into a type-0 quasi-phase-matched periodically poled lithium niobate (PPLN) ridge waveguide (\textit{HC Photonics}) maintained at $55\,^{\circ}\mathrm{C}$ for phase matching.

A programmable pulse shaper (\textit{Waveshaper 4000}) defines $14~\mathrm{GHz}$-wide frequency bin pairs $\{{A}_0,{B}_0\}$ and $\{{A}_1,{B}_1\}$ symmetrically positioned at $\pm112.5~\mathrm{GHz}$ about $\frac{\omega_p-\Delta\omega}{2}$ and $\frac{\omega_p+\Delta\omega}{2}$, respectively.
The logical bin spacing in the Alice-Bob (signal-idler) subspace is thus set by the RF drive frequency $\Delta\omega$. Sara's phase $\theta_S$ is controlled by an analog phase shifter (\textit{Analog Devices}). As an example, the target state $\ket{\phi^+}$ prepared in this manner yields a measured fidelity of $\braket{\phi^+|\rho|\phi^+}=\mathrm{0.9614(30)}$.

The same pulse shaper subsequently routes the signal bins to Alice and the idler bins to Bob. Alice and Bob each use an EOPM (\textit{EOSpace} and \textit{AdvR}) to implement the measurement in \cref{eq:eopm_gate}, with $\theta_A$ and $\theta_B$ likewise controlled by analog phase shifters.
The output frequency bins are demultiplexed using wavelength selective switches (WSSs; \textit{WaveShaper 1000}) and detected by superconducting nanowire single-photon detectors (SNSPDs; \textit{Quantum Opus})~\footnote{In the present experiment, only two SNSPDs are used, one for Alice and one for Bob, with different output-bin combinations measured sequentially. A full implementation of the protocol would require four SNSPDs (two per party) to detect both measurement outcomes simultaneously.}. 
Detection events are recorded by a time tagger (\textit{Swabian}), from which signal-idler coincidence events are identified to reveal the quantum correlations.  The coincidence window is set to $512~\mathrm{ps}$ (comfortably integrating the entire coincidence peak), with an integration time of $10~\mathrm{s}$ for each signal-idler frequency-bin combination. For the correlated combinations, we obtain a coincidence rate of $\sim$80~Hz and a coincidence-to-accidental ratio (CAR) of $\sim$55. 

The RF signals driving Sara's EOIM and the EOPMs used by Alice and Bob are derived from the same signal generator (\textit{Hittite}), providing a common frequency and phase reference across the three parties. For the fast-switching demonstration, arbitrary waveform generators (AWGs) produce electrical control signals at a symbol rate of $20~\mathrm{MHz}$ to drive the analog phase shifters for all three parties. This emulates the operation of the actual protocol, in which the three parties rapidly change their prepared states or measurement bases in real time. Details of the AWGs and RF phase control are provided in Appendix~\ref{app:AWG_PS}, while \cref{sec:dis} mentions pathways for extending to nonlocal scenarios.
The system optical loss is dominated by component insertion loss, including approximately $4.5~\mathrm{dB}$ from each of the three programmable filters (\textit{WaveShapers}) and $4~\mathrm{dB}$ from each of the two EOPMs.

\begin{figure*}[t]
    \centering    \includegraphics[width=\textwidth]{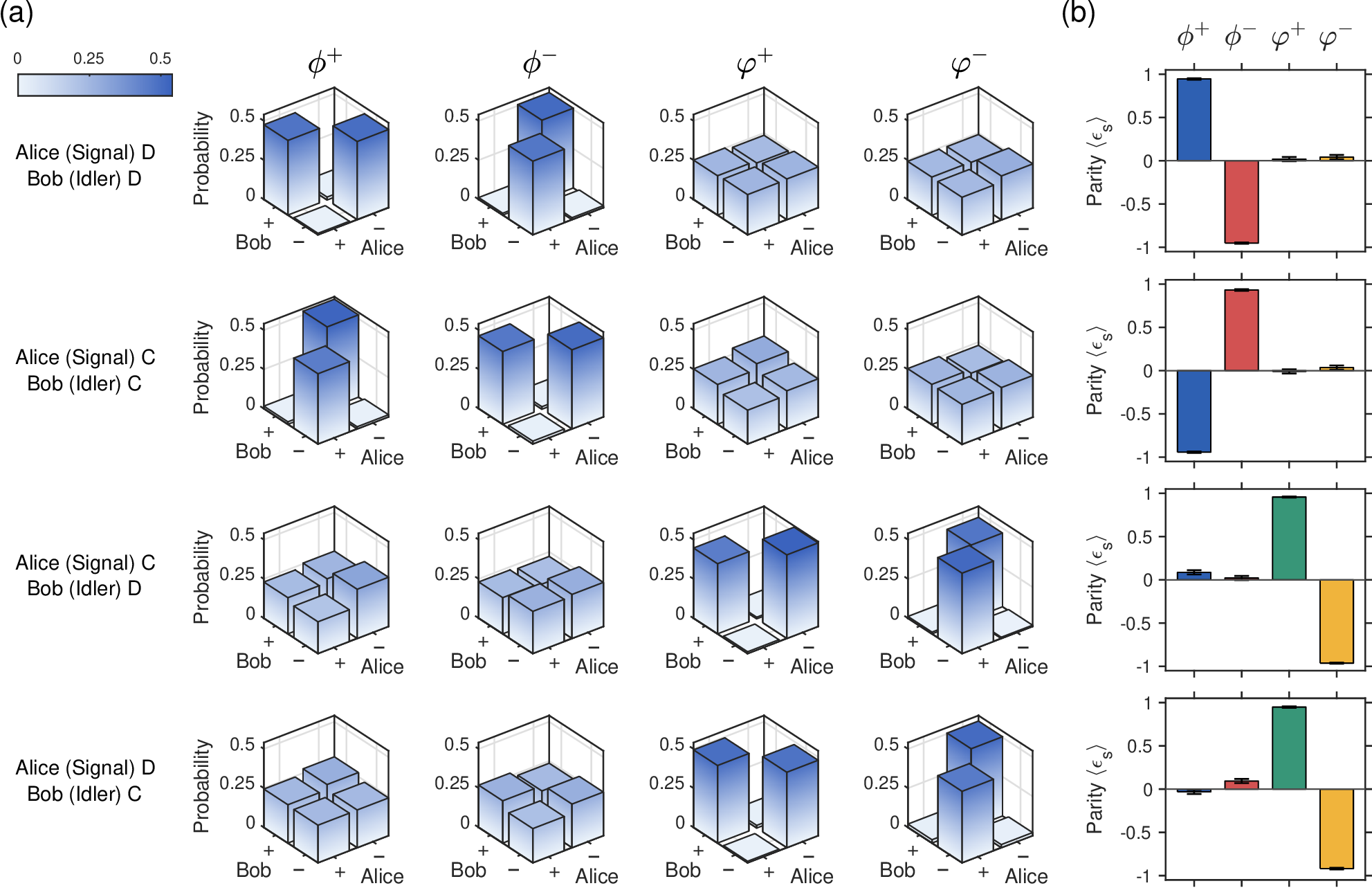}
    \caption{
    Alice-Bob joint outcome probabilities and parity parameters under static configurations. (a) Sixteen cases comprising the four Alice-Bob basis combinations $DD$, $CC$, $CD$, and $DC$ (rows) and Sara's four prepared states $\phi^{+}$, $\phi^{-}$, $\varphi^{+}$, and $\varphi^{-}$ (columns). Each histogram shows the posterior mean probabilities of the four joint outcomes $(+,+)$, $(+,-)$, $(-,+)$, and $(-,-)$, inferred from the measured coincidence counts using a Dirichlet posterior with a uniform prior. (b) Corresponding parity parameters $\braket{\epsilon_s}$ calculated from each row of histograms in (a). Error bars denote $\pm 1$ posterior standard deviation.}
    \label{fig:corr}
\end{figure*}
\section{\label{sec:Results}Results}

\subsection{Static three-party correlations}
To characterize the intrinsic performance of our system without transient errors introduced by active switching, we measure and analyze the joint spectral intensities (JSIs) for all combinations of $(\mathbf{S^s},\mathbf{A},\mathbf{B})$ under static configurations. For these static measurements, the sixteen configurations---Sara's four {states times two measurement bases for each of Alice and Bob}---are measured sequentially. In each case, the state and measurement bases are selected by applying constant control voltages to the corresponding analog phase shifters, which set the RF phases. The resulting coincidence counts are analyzed by Bayesian inference (Appendix~\ref{app:bayesian}) to estimate the underlying outcome probabilities, shown in Fig.~\ref{fig:corr}(a).

By \cref{eq:correlations}, eight of the sixteen combinations are expected to exhibit strong correlations. These cases correspond to the four panels in the upper left and the four panels in the lower right of Fig.~\ref{fig:corr}(a), where the experimentally reconstructed probability distributions closely reproduce the ideal correlated patterns. The remaining eight combinations are expected to be uncorrelated, yielding uniform probability distributions, as also observed experimentally. These uncorrelated configurations cannot be used for secret sharing; however, by biasing the analog waveforms to dwell primarily on predefined dominant bases, asymmetric state and basis selection can be realized to preferentially hit the correlated configurations and improve protocol efficiency~\cite{xiao2004efficient}.

To further quantify the correlations in each case, we adopt the Alice-Bob parity measure $\epsilon_s = ab$, where the subscript $s$ denotes conditioning on Sara's prepared state~\cite{williams2019quantum}. Its expectation value can be calculated as
\begin{equation}
\langle \epsilon_s \rangle
= \bar p_{++} + \bar p_{--} - \bar p_{+-} - \bar p_{-+},
\label{eq:corr_mean}
\end{equation}
using the probability distribution extracted from Fig.~\ref{fig:corr}(a). The resulting values of $\langle \epsilon_s\rangle$ are shown in Fig.~\ref{fig:corr}(b). 

Pairing the corresponding results for $s=+$ and $s=-$ allows us to evaluate the three-party correlation parameter
\begin{equation}
\langle \epsilon \rangle
=
\frac{1}{2}
\left(
\langle \epsilon_{+} \rangle
-
\langle \epsilon_{-} \rangle
\right),
\label{eq:three-party_corr}
\end{equation}
assuming that Sara selects the two private bits $s\in\{+,-\}$ with equal probability. The corresponding quantum bit error rate (QBER) is calculated as
\begin{equation}
\mathrm{QBER}
=
\frac{1}{2}
\left(
1-\left|\langle \epsilon \rangle\right|
\right).
\label{eq:QBER}
\end{equation}
\Cref{tab:StatCorrelation} summarizes the calculated three-party correlations and corresponding QBERs. In all cases, the QBER values remain well below the $11\%$ threshold for entanglement-based QKD with one-way classical communication~\cite{shor2000simple,koashi2003secure}, establishing a strong experimental baseline for QSS operation with our system.

\begin{table}[t]
\centering
\caption{
Combinations of public bits leading to correlated events under static operation, together with the ideal private-bit correlation $\langle\epsilon\rangle$, experimentally measured correlation, and corresponding pairwise QBER.
}
\label{tab:StatCorrelation}
\renewcommand{\arraystretch}{1.15}

\begin{tabular*}{\columnwidth}{@{\extracolsep{\fill}}lccc}
\hline\hline
\shortstack[l]{Public bits\\$(\mathbf{S},\mathbf{A},\mathbf{B})$}
&
\shortstack{$\langle \epsilon \rangle$\\(Ideal)}
&
\shortstack{$\langle \epsilon \rangle$\\(Experiment)}
&
\shortstack{Pairwise\\QBER (\%)} \\
\hline
$(\phi,D,D)$ & $+1$ & $+0.9485(55)$   & $2.58(27) $ \\
$(\phi,C,C)$ & $-1$ & $-0.9372(60)$   & $3.14(30) $ \\
$(\varphi,C,D)$ & $+1$ & $+0.9601(48)$   & $2.00(24)$ \\
$(\varphi,D,C)$ & $+1$ & $+    0.9335(62)$ & $3.32(31)$ \\
\hline\hline
\end{tabular*}
\end{table}

\subsection{Active fast switching}

\begin{figure*}[t]
    \centering    \includegraphics[width=\textwidth]{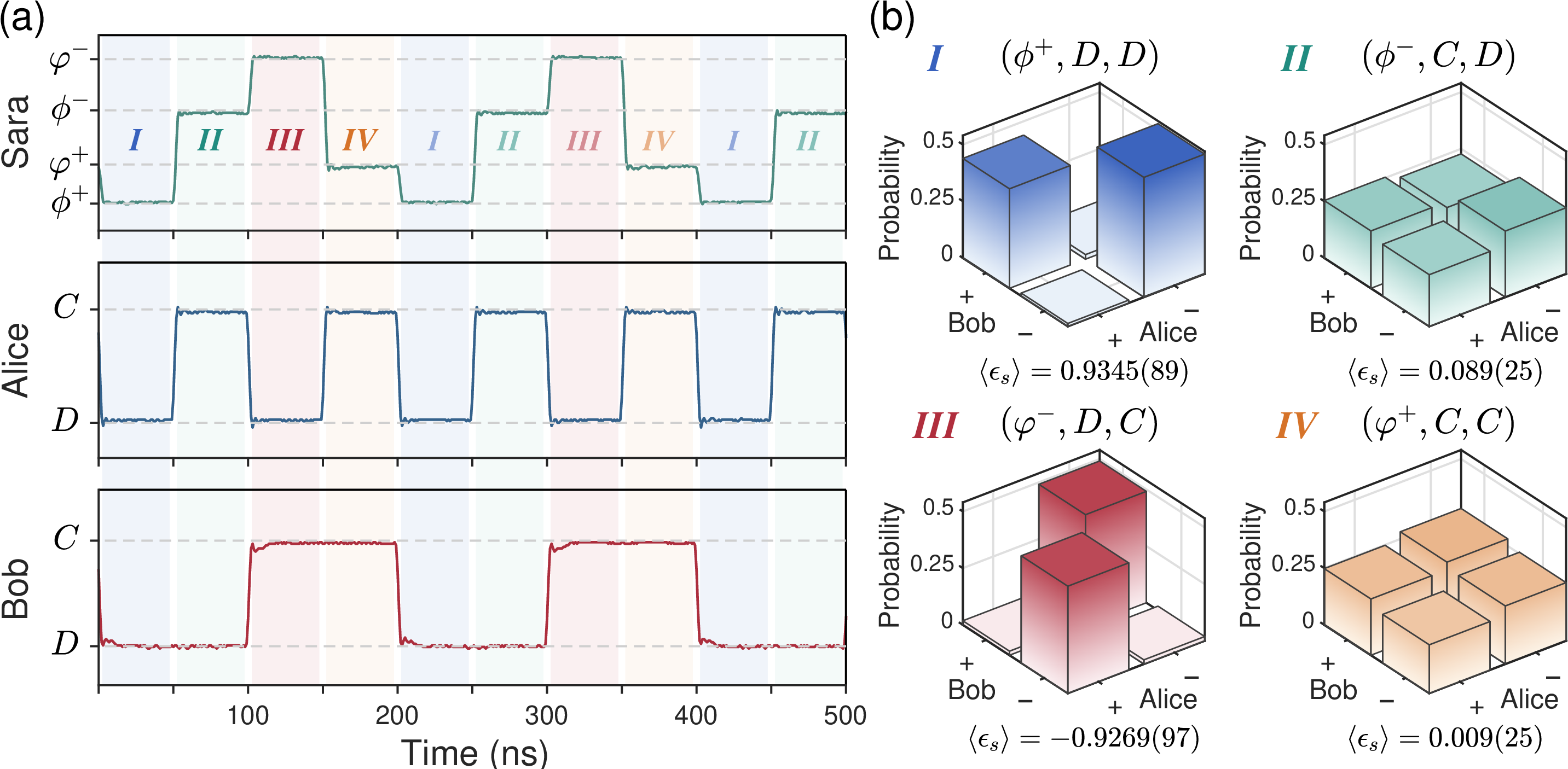}
    \caption{
    Fast active switching among four representative state-basis configurations. (a) Oscilloscope traces of the electrical control signals for Sara's state preparation and Alice's and Bob's measurement-basis selection. Sara's voltage settings to realize the four phases $\theta_S\in \{0,\frac{\pi}{4},\frac{\pi}{2},\frac{3\pi}{4}\}$ are unevenly spaced because of the nonlinear phase-voltage response of the analog phase shifters  (Appendix~\ref{app:AWG_PS}). Each configuration occupies a $50~\mathrm{ns}$ time slot, with the shaded regions marking the central $90\%$ used for analysis. The four configurations $I$--$IV$ repeat periodically. (b) Measured Alice-Bob joint probability distributions for $I=(\phi^{+},D,D)$, $II=(\phi^{-},C,D)$, $III=(\varphi^{-},D,C)$, and $IV=(\varphi^{+},C,C)$. The corresponding values of $\langle\epsilon_s\rangle$ are shown below each distribution, with uncertainties given in parentheses.}
    \label{fig:4state}
\end{figure*}

As an initial test of fast switching, we rapidly switch through a fixed, periodic sequence of four representative configurations, labeled I--IV in Fig.~\ref{fig:4state}, in which the RF phases controlling Sara's state preparation and Alice's and Bob's measurement bases are updated every 50~ns. The control waveforms in Fig.~\ref{fig:4state}(a) are applied to the analog phase shifters, with each voltage plateau corresponding to a particular RF phase and, consequently, to a specific combination of Sara's prepared state and Alice's and Bob's measurement bases.

Using the internal clock of the time tagger together with a marker output from the AWG, which identifies the beginning of configuration I in each cycle, we assign the continuously recorded detection events  to the appropriate switching intervals. Coincidence events associated with each configuration are then accumulated over successive cycles, yielding the joint probability distributions and corresponding parity parameters $\langle\epsilon_s\rangle$.
All four configurations are thus measured within a single acquisition and analyzed separately.

Because the electrical transitions between adjacent voltage levels have a finite duration of approximately $4~\mathrm{ns}$, limited by the bandwidth of the AWGs used in our experiment, only the central $45~\mathrm{ns}$ of each $50~\mathrm{ns}$ switching interval is included in the coincidence analysis, leaving a $2.5~\mathrm{ns}$ guard interval at each edge. This temporal windowing retains $90\%$ of the available detection time, corresponding to an effective loss of approximately $0.5~\mathrm{dB}$. A detailed comparison of the results with and without temporal windowing is provided in Appendix~\ref{app:no_window}.

With temporal windowing and appropriate relative delays applied to the electrical signals driving the three analog phase shifters to ensure that their switching events are well aligned, we obtain the correlations shown in Fig.~\ref{fig:4state}(b). For the two correlated configurations (I and III), the measured values differ by $\sim$1\% from those obtained under static operation [Fig.~\ref{fig:corr}(b)]. This negligible degradation provides a preliminary assessment of the active fast-switching scheme.

However, deterministic switching among only a few predefined configurations does not by itself constitute a protocol-level verification, since a practical implementation requires the prepared states and measurement bases to vary \textit{randomly} from one time slot to another. In this respect, pseudorandom testing plays a role similar to that in classical communication systems, where it is commonly used to sample a broader range of temporal transition patterns and to expose pattern-dependent errors that may remain hidden under periodic operation, including those associated with finite bandwidth, imperfect settling, and timing misalignment.

\begin{figure*}[t]
    \centering    \includegraphics[width=\textwidth]{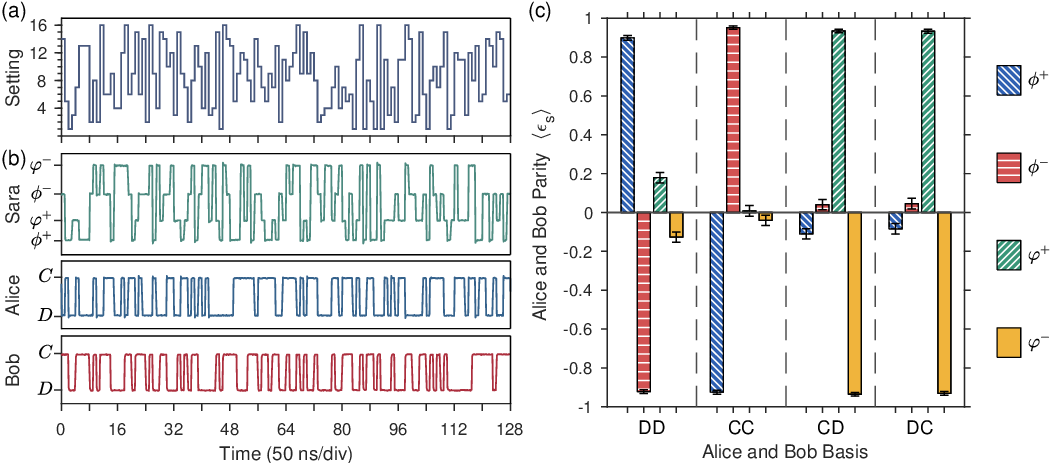}
    \caption{Pseudorandom switching across all state and basis combinations. (a) 128-point pseudorandom pattern used to switch among the 16 state-basis settings, with each switching interval lasting $50~\mathrm{ns}$.
    (b) Corresponding RF phase-control waveforms of Sara, Alice, and Bob, measured on an oscilloscope, for the sequence in (a). (c) Alice-Bob parity $\langle\epsilon_s\rangle$ for all sixteen cases under pseudorandom switching, grouped by the Alice-Bob basis combination (DD, CC, CD, DC); bar patterns denote Sara's state.}    
    \label{fig:rand}
\end{figure*}

In the same spirit, we label all possible state-and-basis configurations outlined in Fig.~\ref{fig:corr} as 1--16 and generate a length-128 pseudorandom sequence, with each element randomly selected from these 16 labels, as illustrated in Fig.~\ref{fig:rand}(a). This sequence is then translated into the corresponding electrical waveforms and applied to the three analog phase shifters [Fig.~\ref{fig:rand}(b)]. As in the preceding measurement, the configuration is updated every $50~\mathrm{ns}$, while the 128-element sequence is repeated continuously throughout the acquisition. Detection events from time slots carrying the same label are grouped together and accumulated over successive repetitions of the sequence. In this way, the correlations for all 16 configurations are obtained separately within a single measurement. The results are shown in Fig.~\ref{fig:rand}(c), with the corresponding three-party correlation parameters summarized in Table~\ref{tab:RandCorrelation}.

\begin{table}[t]
\centering
\caption{
Combinations of public bits leading to correlated events under pseudorandom fast switching, together with the ideal private-bit correlation $\langle\epsilon\rangle$, experimentally measured correlation, and corresponding pairwise QBER.
}
\label{tab:RandCorrelation}
\renewcommand{\arraystretch}{1.15}

\begin{tabular*}{\columnwidth}{@{\extracolsep{\fill}}lccc}
\hline\hline
\shortstack[l]{Public bits\\$(\mathbf{S},\mathbf{A},\mathbf{B})$}
&
\shortstack{$\langle \epsilon \rangle$\\(Ideal)}
&
\shortstack{$\langle \epsilon \rangle$\\(Experiment)}
&
\shortstack{Pairwise\\QBER (\%)} \\
\hline
$(\phi,D,D)$ & $+1$ & $+0.91{05}(7{8})$   & $4.4{7}(39) $ \\
$(\phi,C,C)$ & $-1$ & $-0.93{84}(6{5})$   & $3.{08}(33) $ \\
$(\varphi,C,D)$ & $+1$ & $+0.93{49}(6{8})$   & $3.26(34)$ \\
$(\varphi,D,C)$ & $+1$ & $+    0.93{22}(68)$ & $3.{39}(34)$ \\
\hline\hline
\end{tabular*}
\end{table}

Compared with the static measurements, the pseudorandom fast-switching results exhibit a discernible yet small reduction in correlation below 0.04 for all configurations. This degradation is more pronounced 
than that observed in the four-configuration switching test, consistent with the broader variety of transition patterns involved. The reduction may arise from the finite impulse response of the analog phase shifters, overshoot or waveform distortion at the AWG-generated switching edges, and residual timing misalignment among the electrical signals driving the three parties. Nevertheless, the resulting QBERs remain below $5\%$ for all configurations and well below the $11\%$ security threshold~\cite{shor2000simple,koashi2003secure,ma2007quantum,williams2019quantum}, indicating that high-speed pseudorandom switching can be implemented without substantially compromising system performance.

\section{\label{sec:dis}Discussion}
We have demonstrated the first three-party quantum communication primitive encoded entirely in the frequency-bin degree of freedom, with active electrical control of both state preparation and measurement bases. The strong three-party correlations are preserved under 50~ns-scale fast switching, with QBERs remaining well below the security threshold.

Notably, the same switching techniques can be adapted to frequency-bin QKD by simply turning off the active switching at Sara. In this simplified setting, fast electrical basis switching can reduce hardware overhead while providing the flexibility required for schemes such as biased-basis selection. 
Although the switching time can be increased further with higher-bandwidth phase shifters, the practical benefit of doing so is bounded by the detectors. 
Even in the limit of zero loss, discrete-variable quantum communications is ultimately limited by detector dead time: one cannot process photons faster than the single-photon detectors can receive them~\cite{rogers2007detector}, and basis selection needs to be updated only after each detection event~\cite{Sasaki2016}. 
Our SNSPD dead time of $\sim$25~ns is already nearly in reach of our 50~ns symbol period, while the fastest commercially available, dead-time-optimized SNSPDs (\textit{Single Quantum}) have reached only $\sim$4~ns recovery times~\cite{SingleQuantumInterleaved}.

Looking ahead, an important step toward deployed frequency-bin QSS is to extend the present experiment to physically remote nodes connected by km-scale fiber links. In the current setup, the three parties share a common RF source and are connected by $\sim$10~m-scale fibers. For longer links---like the 26~km spool in  Ref.~\cite{tagliavacche2025frequency}---thermally induced delay fluctuations can cause the EOPMs at Alice and Bob to drift out of sync and increase the QBER. Although the drift can be corrected by relatively slow (few-second-scale) feedback~\cite{tagliavacche2025frequency}, we anticipate this issue can be avoided altogether by distributing RF references through RF-over-fiber (RFoF) links \emph{multiplexed} with the quantum signals. 
In this configuration, the thermal drift is common to both quantum and classical fields, so first-order delay variations have no impact on the quantum correlations. In fact, in a recent experiment exploring nonlocal frequency-bin modulation~\cite{chapman2025quantum}, the delay between the RFoF clock and a multiplexed test signal was found to shift less than 0.5~ps over 30~min in a 5.5~km spool. Auspiciously, then, RFoF techniques for clock distribution should not only expand the possible distances for frequency-bin quantum communications, but also reduce the need for active RF phase tracking.

Beyond long-distance scaling, several improvements could further enhance the performance and scalability of the system, by integrating the advances described here with the directions pioneered in other recent frequency-bin QKD experiments. The source brightness and coincidence window can be optimized to maximize the key rate~\cite{neumann2021model},
and the protocol could be extended to higher-dimensional encoding~\cite{crisan2026multi}. Parallel use of multiple signal-idler channel pairs across the broadband SPDC spectrum, together with frequency-to-time mapping~\cite{khodadad2025frequency}, could further increase the overall communication rate without requiring a proportional increase in detector resources. From a purely hardware perspective, expanded on-chip integration could likewise reduce the system footprint and interconnect loss~\cite{tagliavacche2025frequency}.

Finally, a complete operational QSS system would incorporate raw-key generation, basis sifting, error correction, and subsequent security processing. Combining these protocol-level functions with long-distance synchronization, phase stabilization, and true-random switching driven by a quantum random number generator would move frequency-bin QSS to operation under complete network conditions.

\begin{acknowledgments}
The authors thank AdvR for lending an electro-optic phase modulator. The authors also acknowledge the use of ChatGPT for assistance with language editing and stylistic refinement of the manuscript. Funding of this work was provided by the U.S. Department of Energy (ERKJ381, DE-SC0026396).
\end{acknowledgments}

\appendix

\section{Detailed controls of analog phase shifters}
\label{app:AWG_PS}
\begin{figure}[!tb]
    \centering    \includegraphics[width=\columnwidth]{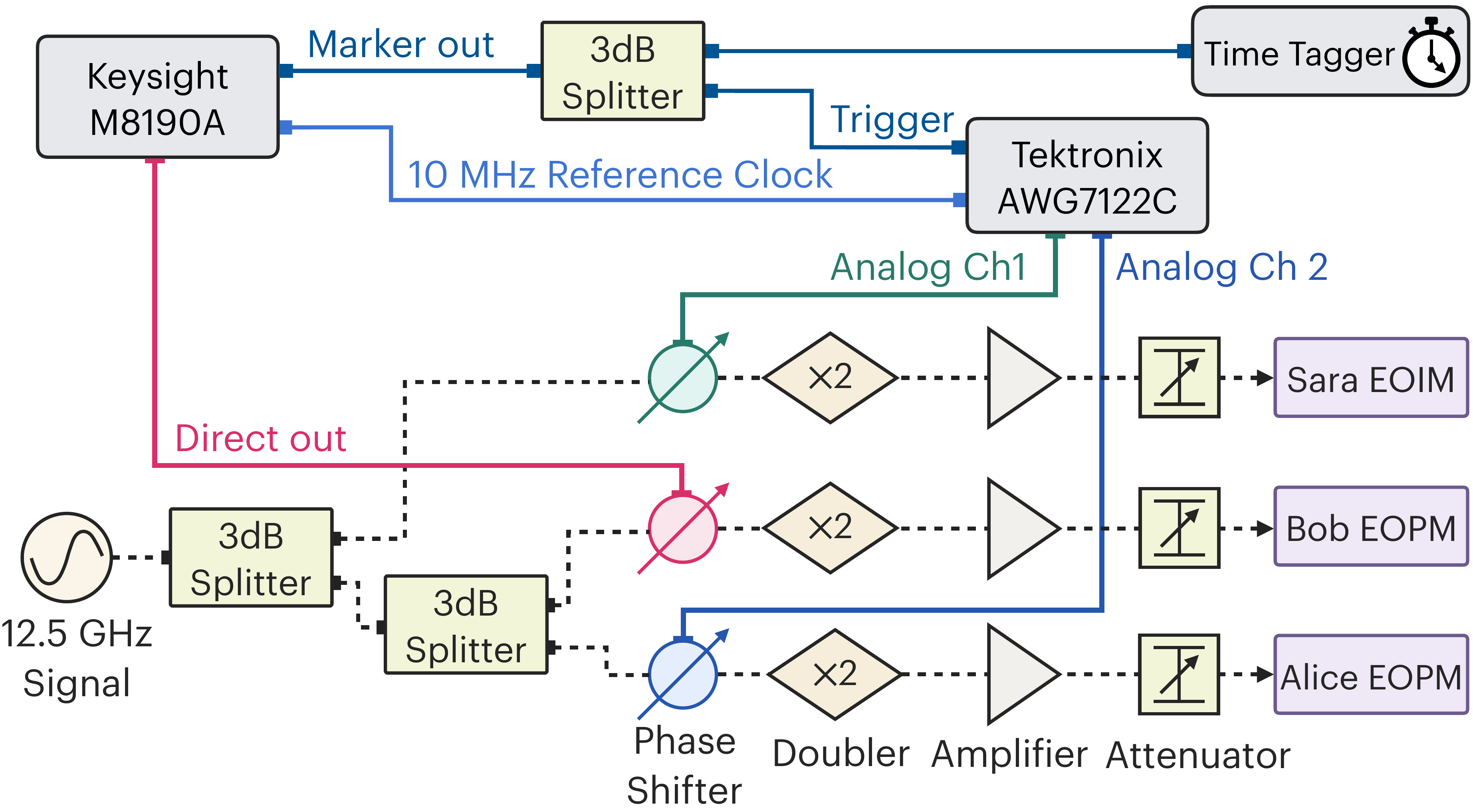}
    \caption{
    RF phase-control and AWG synchronization. Solid lines indicate the paths of MHz-scale electrical control signals, while dashed lines denote GHz-scale paths. A common $12.5~\mathrm{GHz}$ source is phase shifted, frequency doubled, amplified, and attenuated before driving the modulators at $25~\mathrm{GHz}$. The two AWGs (\textit{Keysight M8190A} and \textit{Tektronix AWG7122C}) are locked to a common $10~\mathrm{MHz}$ reference and synchronized by a marker output. A copy of the marker is sent to the time tagger to define the start of each switching pattern.
    }
\label{fig:AWG_PS}
\end{figure}

Figure~\ref{fig:AWG_PS} shows the RF connections used for phase control, together with the synchronization and routing of the AWG signals. Because the analog phase shifters (\textit{ADI HMC-C0101}) operate over 6--15~GHz, a common signal generator (\textit{Hittite HMC-T2100}) first produces a $12.5~\mathrm{GHz}$ tone, whose phase is tuned by the corresponding phase shifter and then frequency doubled to $25~\mathrm{GHz}$. Since we found that tuning the phase shifters also introduces small changes in the RF amplitude, the amplifiers following the doublers are driven deep into saturation, making their output power insensitive to these variations. Variable attenuators are then used to set the RF power, and hence the desired modulation index, at each EOPM or EOIM. This arrangement effectively eliminates changes in modulation depth while the RF phase is varied.

We first sweep the bias voltage of each analog RF phase shifter and monitor the resulting changes in the measured correlations to identify the corresponding RF phase settings. Fast switching is then implemented between the bias voltage levels that yield the optimal results in Fig.~\ref{fig:corr}. The electrical control signals for the three parties are generated using two AWGs (\textit{Keysight M8190A} and \textit{Tektronix AWG7122C})~\footnote{Two AWGs are used because three independent control waveforms are required, exceeding the outputs available from a single unit in our setup.}. The former controls Bob, while the latter controls Alice and Sara. The two AWGs are frequency-locked to a common \(10~\mathrm{MHz}\) reference, and their waveform playback is synchronized using the marker output of the \textit{Keysight}. A marker pulse generated at the beginning of each pseudorandom pattern is used to trigger the \textit{Tektronix} at startup, while a copy is sent to the time tagger to identify the pattern boundary. After compensating for the relative propagation delay, the marker defines the starting point for partitioning the detection record, with the switching-interval duration set by the time tagger's internal clock. The relative delays among the three electrical waveforms are also adjusted to align the three-party switching events with one another and with the data-acquisition windows.

\section{Bayesian estimation of probabilities and correlations}
\label{app:bayesian}
We estimate the outcome probabilities and their statistical uncertainties from the measured coincidence counts using Bayesian inference, focusing on raw probabilities in lieu of quantum state estimation (which imposes additional physical constraints that can artificially reduce QBER). For a given measurement setting and value of Sara's private bit $s$, the coincidence counts $\mathbf{{n}}=({n}_{++},{n}_{+-},{n}_{-+},{n}_{--})$ measured in Alice-Bob outcomes $a,b\in\{+,-\}$ follow a multinomial likelihood
\begin{equation}
\mathcal{L}(\mathbf{{n}}|\mathbf{p})\propto\prod_{a,b}p_{ab}^{\,{n}_{ab}},
\end{equation}
where $\sum_{a,b} p_{ab}=1$ and $\sum_{a,b} {n}_{ab}=N$. Bayes' theorem defines the posterior distribution for $\textbf{p}=(p_{++},p_{+-},p_{-+},p_{--})$ as
\begin{equation}
\pi(\mathbf{p}|\mathbf{{n}})
\propto
\mathcal{L}(\mathbf{{n}}|\mathbf{p})\,\pi_{0}(\mathbf{p}),
\end{equation}
where we take the prior to be the flat Dirichlet density,
\begin{equation}
\pi_{0}(\mathbf{p})=\mathrm{Dirichlet}(\mathbf{p};\boldsymbol{\alpha}_{0}),
\quad\boldsymbol{\alpha}_{0}=(1,1,1,1),
\end{equation}
with $\mathrm{Dirichlet}(\mathbf{p};\boldsymbol{\alpha})\propto\prod_{a,b}p_{ab}^{\,\alpha_{ab}-1}$. For the particular choice of $\boldsymbol{\alpha}_{0}$ here, this density is constant over the probability simplex and therefore assigns equal prior weight to all outcome distributions. Multiplying the multinomial likelihood by this prior gives the posterior density
\begin{equation}
\label{eq:posterior}
\pi(\mathbf{p}|\mathbf{{n}})=\mathrm{Dirichlet}(\mathbf{p};\mathbf{{n}}+1).
\end{equation}
which is again Dirichlet. The mean and standard deviation of the correlation parameter can be computed analytically from the measured counts as
\begin{equation}
\langle\epsilon_{s}\rangle
=\frac{{n}_{++}+{n}_{--}-{n}_{+-}-{n}_{-+}}{N+4}.
\label{eq:bayesian_corr}
\end{equation}
and
\begin{equation}
\sigma_{\epsilon_{s}}
=\sqrt{\frac{1-{\langle\epsilon_{s}\rangle}^{\,2}}{N+5}}.
\label{eq:bayesian_corr_std}
\end{equation}
Similarly, assuming statistically independent coincidence measurements for the two values of \(s\), the uncertainties in the three-party correlation parameter [Eq.~(\ref{eq:three-party_corr})] and the QBER [Eq.~(\ref{eq:QBER})] are
\begin{equation}
\sigma_{\epsilon}
=\frac{1}{2}\sqrt{\sigma_{\epsilon_{+}}^{2}+\sigma_{\epsilon_{-}}^{2}},
\qquad
\sigma_{\mathrm{QBER}}=\frac{1}{2}\,\sigma_{\epsilon}.
\end{equation}

\section{Pseudorandom test without temporal windowing}
\label{app:no_window}
\vspace{-4pt}
\begin{table}[!b]
\vspace{-8pt}
\caption{Measured three-party correlation parameter for the correlated combinations under pseudorandom switching without temporal windowing. The fifth column lists the QBER increment relative to the temporally windowed results in Table~II.}
\label{tab:no_window}
\centering
\begin{tabular*}{\columnwidth}{@{\extracolsep{\fill}}ccccc@{}}
\hline\hline
Public bits
& $\langle\epsilon\rangle$
& $\langle\epsilon\rangle$
& Pairwise
& $\Delta$QBER \\
$(\mathbf{S},\mathbf{A},\mathbf{B})$
& (Ideal)
& (Experiment)
& QBER (\%)
& (\%) \\
\hline
$(\phi,D,D)$       & $+1$ & $+0.8{734}(9{0})$ & ${6.33}(4{5})$ & $+{1.86}$ \\
$(\phi,C,C)$       & $-1$ & $-0.{8955}({80})$ & ${5.23}({40})$ & $+{2.15}$ \\
$(\varphi,C,D)$    & $+1$ & $+0.9{060}(7{8})$ & ${4.70}(3{9})$ & $+{1.44}$ \\
$(\varphi,D,C)$    & $+1$ & $+0.9{216}(7{2})$ & ${3.92}(3{6})$ & $+{0.53}$ \\
\hline\hline
\end{tabular*}
\end{table}

To characterize the system performance without temporal windowing, we repeat the measurement in Fig.~\ref{fig:rand} without applying temporal guard windows around the switching transitions. Table~\ref{tab:no_window} summarizes the resulting three-party correlations and pairwise QBERs. Compared with the temporally windowed results in Table~\ref{tab:RandCorrelation}, the QBER increases by 0.53--2.{15}\%, with the largest value reaching $\sim${6.4}\%---still comfortably below the 11\% security threshold, suggesting that temporal windowing is not strictly required in the present setup.

\bibliography{apssamp}

\end{document}